\documentclass[twocolumn,aps,showpacs,preprintnumbers,amsmath,amssymb,ltxgrid]{revtex4}

\usepackage{graphicx,color,epsfig}

\begin{document}

\bibliographystyle{apsrev}

\preprint{version {\today}}

\title{Comment on ``Direct evidence for hidden one-dimensional Fermi surface of hexagonal K$_{0.25}$WO$_3$''}

\author{Roberta Poloni$^{\dag}$}
\author{Enric Canadell$^{\dag}$}
\author{Jean Paul Pouget$^{\ddag}$}

\affiliation{
(\dag) Institut de Ci\`{e}ncia de Materials de Barcelona (CSIC), Campus de la UAB, 08193, Bellaterra, Spain}
\affiliation{
(\ddag) Laboratoire de Physisque des Solides, CNRS UMR 8502, Universit\'{e} de Paris-Sud, 91405 Orsay, France}

\date{\today}

\begin{abstract}
\end{abstract}

\maketitle

Low-dimensional molybdenum and tungsten oxides and bronzes have been the focus of much attention because of the charge density wave (CDW) instabilities they exhibit \cite{book,book2}. 
An important step in our present understanding of the physics of low-dimensional metals was the discovery of non linear transport due to sliding of the CDW in the K$_{0.3}$MoO$_3$ blue bronze \cite{Dum1983}.
 This report launched a wide interest on these materials
and, as a result, a number of quite intriguing discoveries related to the condensation of CDWs were reported \cite{book,book2}. The CDW periodicity is incommensurate with the lattice constant in the blue bronzes or the $\eta-$ and $\gamma-$Mo$_4$O$_{11}$ Magn\'eli phases but commensurate in the sodium and potassium purple bronzes. Another widely studied family of these compounds is that of the monophosphate tungsten bronzes, which exhibit successive CDWs with modulations varying considerably for the different members of the family \cite{book2,Gre1993}. The condensation of these CDWs is usually driven by instabilities of the Fermi surface (FS) \cite{Can1991}. 
In some cases the FS clearly exhibits nesting features which are at the origin of the CDW instability. However, in other cases the FSs are apparently non nested and for some time it was difficult to understand the origin of the CDWs.
 The development of the so called {\it hidden nesting} concept \cite{Wha1991} provided a simple way for analyzing the apparently non nested FSs in terms of a superposition of nested ones. 

\begin{figure}[ht!]
\begin{centering}
\includegraphics[scale=0.4]{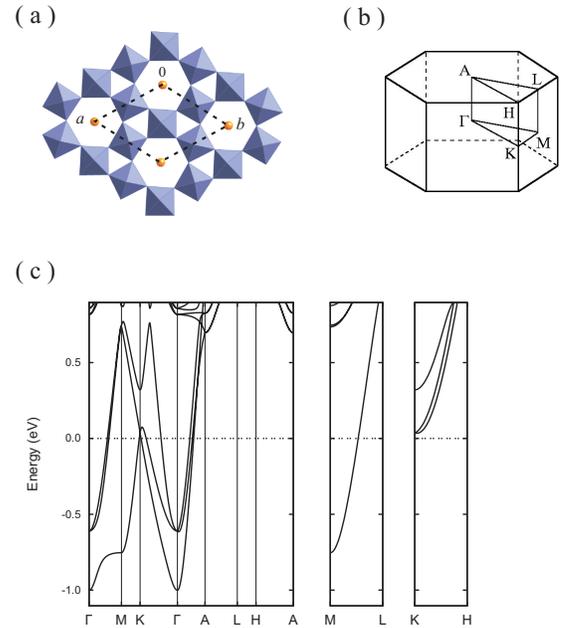}
\caption{(Colour online) (a) Crystal structure of the K$_{0.25}$WO$_3$ system; (b) Hexagonal Brillouin zone of  K$_{0.25}$WO$_3$; (c) Calculated band structure for  K$_{0.25}$WO$_3$ where the dashed line refers to the Fermi level.}
\label{figure1}
\end{centering}
\end{figure}

 Recently, Raj {\it et al.} \cite{Raj2008} have reported angle-resolved photoemission spectroscopy (ARPES) and first-principles density functional theory (DFT) results concerning the FS of the hexagonal potassium tungsten bronze, K$_{0.25}$WO$_3$ \cite{Sch1986} (see Fig.~\ref{figure1}(a)). On the basis of their results they have suggested that the well known high-temperature resistivity anomaly of this phase originates from a possible CDW instability as a result of the existence of hidden pseudo one-dimensional (1D) bands. However, this conclusion is incorrect since it is based on the analysis of data for the $\Gamma$-$K$-$M$ section only of the hexagonal Brillouin zone (see Fig.~\ref{figure1}(b)). Here we report first-principles DFT results showing that K$_{0.25}$WO$_3$ {\it does not possess pseudo-1D partially filled bands} and is instead a 3D metal. In addition, we analyze how the crystal structure determines the nature and shape of the partially filled bands and thus the transport properties. 

   The present calculations were carried out using a numerical atomic orbitals DFT \cite{Hoh1964} approach implemented in the SIESTA code \cite{SolArt2002}. We have used the generalized gradient approximation to the exchange and correlation potential and, in particular, the functional of Wu and Cohen \cite{Wu2006}. Only the valence electrons are considered in the calculation, with the core being replaced by norm-conserving scalar relativistic pseudopotentials \cite{Tro1991} factorized in the Kleinman-Bylander form \cite{Kle1982}. We have used a split-valence double-$\zeta$ basis set including polarization orbitals for all atoms, as obtained with an energy shift of 0.02 Ry \cite{Art1999}. The energy cutoff of the real space integration mesh was 300 Ry and the Brillouin zone (BZ) was sampled using a grid of (12x12x12) $k$-points \cite{Mon1976}.
 The existence of partially empty potassium sites was taken into account via the virtual crystal approximation.

The essence of the reasoning by Raj {\it et al.} \cite{Raj2008} is that both the calculated and ARPES ($a^*$$b^*$) sections of the FS result from the weak hybridization of pairs of parallel lines, perpendicular to the ($a^*$+$b^*$) direction and those equivalent by 120$^\circ$ rotations.
From this observation they conclude that ``the FS is the consequence of hidden (1D) one-dimensional bands''. This clearly means that these bands should have a nil dispersion along the $c^*$ direction. The calculated band structure for K$_{0.25}$WO$_3$ in the region around the Fermi level is shown in Fig. ~\ref{figure1}c. It is immediately evident that the three partially filled bands exhibit a very large dispersion along $c^*$ (i.e. along $\Gamma$ $\rightarrow$ $A$, $M$ $\rightarrow$ $L$ and $K$ $\rightarrow$ H). Consequently, there are no partially filled 1D bands along the direction ($a^*$+$b^*$) and those equivalent by 120$^\circ$ rotations. Since the three partially filled bands are already slightly above the Fermi level at the $K$ point, this means that their dispersion in the $c^*$=0.0 section is also large and consequently, the partially filled bands of K$_{0.25}$WO$_3$ should be 2D or 3D.

\begin{figure}[ht!]
\begin{centering}
\includegraphics[scale=0.4]{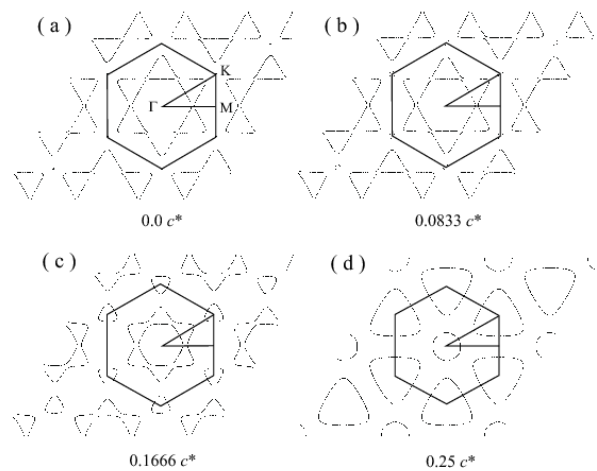}
\caption{Calculated FS for K$_{0.25}$WO$_3$ for different sections of the Brillouin zone perpendicular to the $\Gamma$ $\rightarrow$ $A$ direction (i.e. $c^*$): (a) 0.0$c^*$, (b) 0.0833$c^*$, (c) 0.1666$c^*$ and (d) 0.25$c^*$.}
\label{figure2}
\end{centering}
\end{figure}

The calculated FS for K$_{0.25}$WO$_3$ is reported in Fig.~\ref{figure2} for different sections perpendicular to the $c^*$ direction. The FS in the plane at $c^*$=0.0 is practically identical with the theoretical and experimental sections reported by Raj {\it et al.} \cite{Raj2008} (the very slight differences are due to the different way in which the potassium doping has been treated in the calculations: via a rigid band approach using hexagonal WO$_3$ by  Raj {\it et al.} \cite{Raj2008} or through the explicit consideration of K cations via a virtual crystal approximation here). This section may indeed be seen as resulting from the interaction of three pairs of parallel lines. However, it is also clear that when departing from the plane at $c^*$=0.0 the shape of the FS section seriously changes. In the plane at $c^*$=0.25 any trace of the apparent hidden 1D bands is already lost and none of the sections at $c^*$ $\geq$ 0.3 contributes to the FS. Thus, the three components of the FS are 3D as expected from the calculated band structure. In fact, this conclusion could have been reached by Raj {\it et al.} \cite{Raj2008} since they noted that there is no band crossing the Fermi level in the $AHL$ plane. Alternatively, simple electron counting arguments also lead to the same conclusion. The K$_{0.25}$WO$_3$ unit cell  contains six formula units so that there are 1.5 electrons to fill the three t$_{2g}$-type bands of Fig.~\ref{figure1}(c). 
Consequently, if three bands were 1D they should be one-quarter filled and the nesting vectors can not be those proposed by  Raj {\it et al.} \cite{Raj2008} which would be appropriate for half-filled 1D bands. 

Coming back to the FS section of Fig.~\ref{figure2}, let us note that flat portions with weak curvature may remain until sections with $c^*\sim$ 0.1. Therefore, it can not be completely discarded that weak maxima of the electron-hole Lindhard response function can arise. However, even assuming that this weak feature could stabilize a CDW modulation \cite{book3}, the wave vectors of the modulations proposed by Raj {\it et al.} \cite{Raj2008} are not those which would be expected. 
In fact, when a FS like that of Fig.~\ref{figure2}(a) results from the weak interaction of 1D bands, the appropriate nesting vectors, i.e. those leading to strong maxima in the Lindhard function, are those which allow the simultaneous nesting of two different pairs of parallel lines. This is the so called {\it hidden nesting} \cite{Wha1991} mechanism underlying the CDW instabilities exhibited by a number of oxides and bronzes like the potassium and sodium purple bronzes, the   $\eta-$ and $\gamma-$Mo$_4$O$_{11}$ Magn\'eli phases and several monophosphate tungsten bronzes \cite{book,book2,Gre1993,Can1991}. In the present case, the nesting vectors would be 1/2$a^*$, 1/2$b^*$ and (1/2$a^*$-1/2$b^*$). As mentioned above, although the condensation of CDW modulations of this type can not be completely discarded, the driving force should be quite weak. Let us mention that the superlattice spots reported by Krause {\it et al.} \cite{Kra1985} are quite sample dependent. Moreover, although the nesting vectors proposed by Raj {\it et al.} \cite{Raj2008} may be found in the diffraction patterns, other wave vectors also exist  \cite{Kra1985}. In view of these observations we believe that there is no compelling reason to believe that the superlattice spots and the weak resistivity cusp at 350 K \cite{Kra1985} are due to the condensation of CDWs. We find the arguments of Krause {\it et al.} \cite{Kra1985} concerning the ordering and migration of potassium vacancies more convincing. Also, the FS determined by photoemission at 20 K in the ($a^*$$b^*$) reciprocal plane \cite{Raj2008} does not reveal the formation of partial gaps expected from the stabilization of a CDW.

\begin{figure}[h!]
\begin{centering}
\includegraphics[scale=.5]{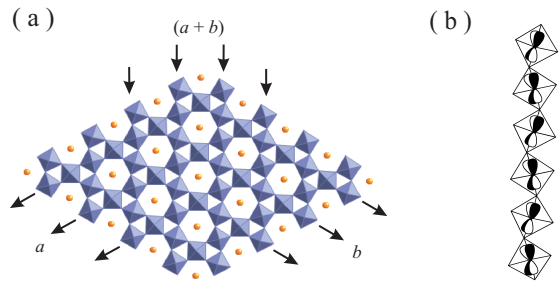}
\caption{(Colour online) (a) Schematic diagram showing that every hexagonal octahedral plane can be seen as resulting from the intersection of octahedral chains along the $a$, $b$ and  ($a$+$b$) directions. (b) Top view of a $xz/yz$ orbitals chain along the ($a$+$b$) direction (see text).}
\label{figure3}
\end{centering}
\end{figure}

With this analysis in mind it is puzzling that both the experimental and theoretical section of the FS for $c^*$=0.0 is made of sets of weakly interacting pairs of parallel lines, as if it was originating from partially filled 1D bands. This feature is however easy to understand after analysis of the band structure \cite{Can1991}. The three partially filled bands are essentially built from the tungsten $xz$ and $yz$ orbitals which mix with the oxygen $p$ orbitals in an antibonding way. In an ideal hexagonal structure, the tungsten $xz$ and $yz$ orbitals are antisymmetric with respect to the basal planes of the octahedra. For wave functions corresponding to a 0.0 $c^*$ component of the $k$-vector, the $p_{x,y}$ orbitals of the apical oxygens can not mix by symmetry into the wave function. Consequently, there is no interaction between the tungsten $xz/yz$ orbitals of different hexagonal planes for this specific section of the Brillouin zone. In this particular case, the band dispersion can only arise from interactions within a hexagonal plane. The pseudo 1D nature of the bands for this section is easy to understand when it is realized that: (a) as shown in Fig.~\ref{figure3}(a), one hexagonal octahedral plane can be seen as resulting from the intersection of octahedral chains along the $a$, $b$ and ($a$+$b$) directions, in such a way that every octahedron belongs simultaneously to two different types of chains, and (b) every octahedron possesses one pair of $xz/yz$ orbitals which can mix and rotate leading to two new orbitals avoiding as much as possible the antibonding mixing with the basal oxygen p$_z$ orbitals, thus leading to the lowest energy t$_{2g}$-based crystal orbitals. In other words, every octahedron can use independently the two appropriate $xz/yz$ combinations to make extended interactions along the two chains common to every octahedron. This leads to chains of orbital interactions like that schematically shown in Fig.~\ref{figure3}(b) (top view), which run along the  ($a$+$b$) direction. These chains of orbitals do not interact with orbitals of parallel chains because the tungsten $xz/yz$ combination can not overlap with the $p_z$ orbital of the oxygen atoms placed at the basal site perpendicular to the chain direction as a result of the local symmetry (see Fig.~\ref{figure3}b). Hence, the absence of such oxygen contributions isolates different parallel chains. In addition, the fact that two types of chains intersect at each octahedron does not lead to a sizeable interaction because of the local orthogonality of the two combinations of $xz/yz$ orbitals. Taking into account the actual distortions of the lattice with respect to the ideal one does not alter in any significant way this description of the partially-filled bands. As a matter of fact, the crystal orbitals at $\Gamma$ are essentially made of sets of parallel chains of orbitals like that of Fig.~\ref{figure3}(b) (or those equivalent along the  $a$ and $b$ directions) which practically do not communicate. When the component of the $k$-vector changes along the direction of the chain, the $p_z$ orbitals of the oxygens shared along the chain can mix into the crystal orbital because of the associated phase changes and the band acquires dispersion. However, when the $k$-vector component changes along the interchain direction, the chains practically do not communicate because of the local symmetry argument pointed above. This means that for this particular section of the Brillouin zone the system can be considered to result from the very weak hybridization brought about by the octahedral distortions of three 1D chains running along the $a$, $b$ and ($a$+$b$) directions. However, as soon as the $c^*$ component of the $k$-vector departs from zero the inter-octahedral planes interactions are switched on and any trace of the pseudo-1D behavior is lost.  
In summary, we find that K$_{0.25}$WO$_3$ does not exhibit 1D partially filled bands and that even if the Lindhard function could exhibit weak maxima due to nesting of the FS, the more intense ones should not be those previously proposed \cite{Raj2008}. We believe that arguments based on potassium vacancy ordering are more likely when trying to explain the origin of the resistivity anomaly of this bronze.

Work at Bellaterra was supported by DGI-Spain (Projects FIS2006-12117-C04-01 and CSD2007-00041), Generalitat de Catalunya (2005 SGR 683) and by grants for computer time from the CESCA and CESGA.

\onecolumngrid

\end{document}